\documentclass[letterpaper]{article} 
\usepackage{aaai25}  

\usepackage{times}  
\usepackage{helvet}  
\usepackage{courier}  
\usepackage[hyphens]{url}  
\usepackage{graphicx} 
\usepackage{natbib}  
\usepackage{caption} 
\usepackage{amsmath}
\usepackage{amssymb}
\usepackage{algorithm}
\usepackage{algorithmic}
\usepackage{booktabs}
\usepackage{newfloat}
\usepackage{listings}
\DeclareCaptionStyle{ruled}{labelfont=normalfont,labelsep=colon,strut=off} 
\floatstyle{ruled}
\newfloat{listing}{tb}{lst}{}
\floatname{listing}{Listing}
\title{ProteinRPN: Towards Accurate Protein Function Prediction with Graph-Based Region Proposals}
\author{Shania Mitra\equalcontrib\textsuperscript{\rm 1},
    Lei Huang\equalcontrib\textsuperscript{\rm 1, \rm 2},
    Manolis Kellis\textsuperscript{\rm 1}
}
\affiliations{
    \textsuperscript{\rm 1} Massachusetts Institute of Technology    \textsuperscript{\rm 2}City University of Hong Kong \\

shania@mit.edu, lhuang93-c@my.cityu.edu.hk, manoli@mit.edu
}

\usepackage{bibentry}

\begin{document}

\maketitle

\begin{abstract}

Protein function prediction is a crucial task in bioinformatics, with significant implications for understanding biological processes and disease mechanisms. While the relationship between sequence and function has been extensively explored, translating protein structure to function continues to present substantial challenges. Various models, particularly, CNN and graph-based deep learning approaches that integrate structural and functional data, have been proposed to address these challenges. However, these methods often fall short in elucidating the functional significance of key residues essential for protein functionality, as they predominantly adopt a retrospective perspective, leading to suboptimal performance. 

Inspired by region proposal networks in computer vision, we introduce the Protein Region Proposal Network (ProteinRPN) for accurate protein function prediction. Specifically, the region proposal module component of ProteinRPN identifies potential functional regions (anchors) which are refined through the hierarchy-aware node drop pooling layer favoring nodes with defined secondary structures and spatial proximity. The representations of the predicted functional nodes are enriched using attention mechanisms and subsequently fed into a Graph Multiset Transformer, which is trained with supervised contrastive (SupCon) and InfoNCE losses on perturbed protein structures. Our model demonstrates significant improvements in predicting Gene Ontology (GO) terms, effectively localizing functional residues within protein structures. The proposed framework provides a robust, scalable solution for protein function annotation, advancing the understanding of protein structure-function relationships in computational biology.

\end{abstract}


\section{Introduction}
Advancements in genomics technology have illuminated the study of protein functions, enabling researchers to uncover the roles and interactions of proteins within living systems, making this a pivotal task in modern biology. Despite the vast number of proteins available, only a few of them have been reviewed by human curators. Among these reviewed proteins, less than 19.4\% are substantiated by wet-lab experimental evidence \cite{uniprot2023uniprot}. Precise functional annotations of proteins are crucial for tasks such as pinpointing drug targets, unraveling disease mechanisms, and enhancing biotechnological applications across industries \cite{kulmanov2024protein}.

Currently, Gene Ontology (GO) \cite{gene2023gene, gene2021gene} stands out as the most comprehensive resource, embodying all the essential attributes of an ideal functional classification system. The GO consortium delineates the functional attributes of genomic products, including genes, proteins, and RNA. Specifically, GO utilizes three subontologies to organize function terms according to each product: Biological Process (BP), Molecular Function (MF), and Cellular Component (CC). Although the UniProtKB/Swiss-Prot database records manually curated GO annotations that are verified by wet-lab experiments, there are still a significant number of protein sequences lacking functional annotations due to the high costs and limited throughput of experimental studies.

Fortunately, machine learning methods have emerged as a promising tool to address this challenge. Recently developed machine learning methods leverage different protein information for function prediction, including protein sequential information, protein tertiary structure, protein-protein interaction (PPI) networks, phylogenetic analysis, and literature information \cite{GOLaberler2018, you2021deepgraphgo, you2019netgo, gligorijevic2021structure, lai2022accurate, kulmanov2022deepgozero, kulmanov2018deepgo, pan2023pfresgo, kulmanov2024protein,gu2023hierarchical}. Specifically, early studies focused on learning the similarities of homologous proteins by utilizing sequence alignment tools \cite{gong2016gofdr}. This idea was then extended to harness additional protein information, such as PPI networks and biophysical properties, to predict protein function \cite{cho2016compact, you2021deepgraphgo, you2019netgo, pan2023pfresgo, cho2016compact}. However, the sequential similarity of proteins alone cannot fully determine protein function. Furthermore, these knowledge-based models heavily rely on selected features and cannot be generalized to new proteins due to the absence of prior knowledge.

Subsequent studies leverage primary sequence as the main feature for function prediction \cite{kulmanov2018deepgo, kulmanov2022deepgozero}. While the relationship between sequence and function has been extensively investigated, translating protein structure into function remains a significant challenge. Various models, notably CNNs and graph-based deep learning approaches that incorporate both structural and functional information, have been proposed to tackle these hurdles \cite{gligorijevic2021structure, lai2022accurate, gu2023hierarchical}. However, these methods often fall short in elucidating the functional significance of key residues essential for protein functionality. Most of these approaches employ post-hoc techniques, such as Gradient-based Class Activation Maps \cite{gu2023hierarchical, gligorijevic2021structure}, to provide visual explanations of which residues contribute most to the predicted function. Yet, this retrospective analysis lacks biological insight, as it relies solely on what the model has learned during training without accounting for prior knowledge about functional residues. Moreover, these methods often result in a selection of numerous scattered residues with low specificity, diluting the focus on the truly important regions and leading to suboptimal performance.
To address these limitations, we introduce ProteinRPN, a novel model for accurate protein function prediction. ProteinRPN intentionally incorporates functional residue detection, enabling it to prioritize critical regions where groups of residues work together to perform specific functions. Inspired by region proposal networks in computer vision \cite{ren2015faster,tang2018weakly}, ProteinRPN incorporates a graph-based Region Proposal submodule to identify potential functional regions within proteins. The model starts by detecting regions which contain functional residues, focusing on $k$-hop subgraphs (anchors) surrounding each node. These identified functional regions are then refined harnessing a node drop pooling layer, which prioritizes nodes with defined secondary structures and spatial proximity, employing hierarchy-aware attention to assess functionality. The representations of these functional nodes are further enriched through a functional attention layer. Finally, the Graph Multiset Transformer (GMT) converts node-level representations into comprehensive graph-level embeddings, integrating locally emphasized interactions while preserving the global graph structure. Additionally, we utilize contrastive learning to generate similar representations for functionally related proteins while ensuring that distinct proteins have distinct representations.

The region proposal module is initially pretrained on the PDBSite dataset \cite{Ivanisenko2000PDBSite}, containing functional residue annotations sourced from the Protein Data Bank (PDB) \cite{berman2000protein}, which is a popular database known for its experimentally derived structural data on proteins. Then, we conduct experiments on the same dataset as baseline models \cite{gligorijevic2021structure, you2021deepgraphgo, gu2023hierarchical} for a fair comparison with baselines. The experimental results indicate significant improvements in predicting protein functions compared to state-of-the-art (SOTA) models. Remarkably, the proposed model achieves a $\sim$7\% improvement in protein-centric Fmax on BP and MF ontologies compared to SOTA models. We also visualize the predicted functional residues, demonstrating that our model can identify essential functional structures and regions, which are meaningful for biological analysis.

\section{Related Work}
Computational methods have been proposed for protein function prediction, offering a more efficient and less resource-intensive alternative to wet-lab experimental assays. The task is framed as a multiclass multilabel classification problem, where each protein can be associated with multiple GO terms. Due to the hierarchical structure of GO terms within an ontology, predicting a given term also implies predicting all its ancestor terms, adding an additional layer of complexity. Early studies \cite{tian2004eficaz, gong2016gofdr} leveraged query sequence-based Multiple Sequence Alignments (MSA) to predict protein GO terms. Based on the Position-Specific Scoring Matrix (PSSM), these models could identify query sequences that are more similar to sequences in the homo-functional MSA. Consequently, the protein sequence is more likely to be annotated with the target GO term. 

Machine learning models have since emerged for more accurate protein function prediction by utilizing a broader range of biological features. Some methods \cite{GOLaberler2018, you2019netgo} rely on external knowledge or even the hierarchical structure of GO terms, including GO term frequency, sequence alignment, amino acid trigram, domains and motifs, biophysical properties, and PPI networks. These approaches often employ a learning to rank (LTR) \cite{li2011short} framework for automatic function prediction. Sequence-based methods \cite{fa2018predicting, kulmanov2018deepgo, wang2023netgo} utilize sequential models like 1D CNNs and Transformers to derive protein sequence representations. Given that Graph Neural Networks (GNNs) are well-suited for learning the topology of PPI networks, subsequent studies \cite{zhao2022panda2} have combined hybrid features from protein sequences and PPI networks, embedded using GNN modules, for function prediction.

Since protein structures determine essential biological and chemical properties \cite{challengeCJ2023}, relying exclusively on sequence-based methodologies may present a significant limitation. Therefore, several studies have incorporated protein structures for more accurate predictions \cite{gligorijevic2021structure, lai2022accurate, gu2023hierarchical}. Specifically, these models derive contact maps from protein structures to construct residue graphs. Additionally, as protein amino acid sequences are similar to natural language sentences, recent studies \cite{gu2023hierarchical} utilize advanced protein language models like ESM-1b \cite{rives2021biological} to obtain richer sequence representations. However, there remains a gap in models that accurately detect and predict constellations of amino acids in protein active sites and leverage these for structural and functional insights \cite{challengeCJ2023}.
\section{Methodology}

\begin{figure*}[ht]
    \centering
    \includegraphics[width=0.9\textwidth]{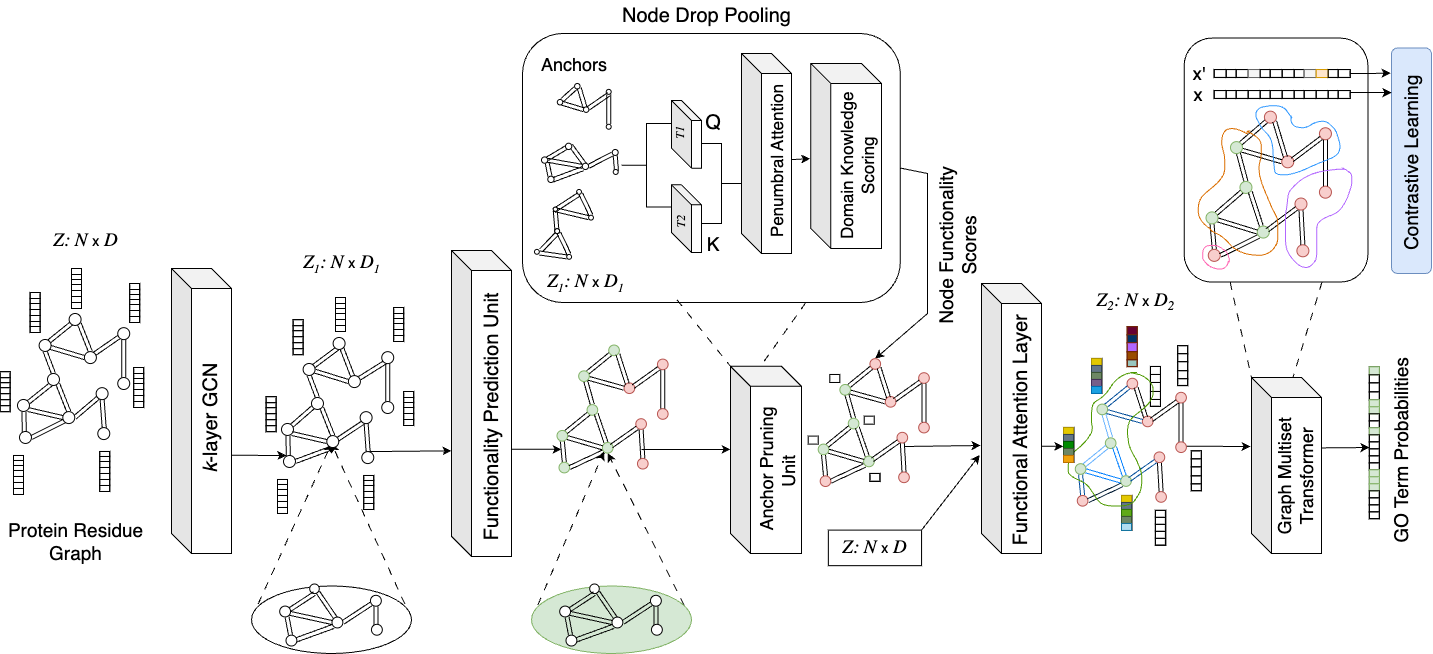}
    \caption{The ProteinRPN model predicts protein function by converting protein sequences into residue graphs, processing them through a k-layer GCN to identify functional subgraphs (anchors), refining these subgraphs via  domain knowledge and hierarchy-aware attention mechanisms, and categorizing them into GO terms using a GMT layer}
\label{fig:model}
\end{figure*}
In this section, we introduce ProteinRPN, a novel model for protein function prediction. As illustrated in Figure \ref{fig:model}, ProteinRPN operates on protein graphs where nodes represent individual residues and edges are defined by the contact map which reflects residue proximity within the three-dimensional structure. The architecture is composed of three primary components. The first component, Region Proposal Network, is responsible for processing the protein graphs and proposing subgraphs which contain functionally relevant regions. These subgraphs are then fed through Functional Attention Layer which enhances the region proposals and selectively amplifies the representations of functional regions through a learned attention mechanism. The refined representations are subsequently passed to the Function Prediction block, consisting of a Graph Multiset Transformer (GMT) pooling layer and an MLP readout layer which generates predictions for GO terms. The entire framework is optimized through a combination of Supervised Contrastive (SupCon) loss and a self-supervised Information Noise-Contrastive Estimation (InfoNCE) loss, ensuring robust and effective protein representation learning. 

\subsubsection{Motivation} Our architecture is motivated by an analysis of 603 protein structures from PDBSite \cite{Ivanisenko2000PDBSite}, which reveals that functional residues tend to cluster in three-dimensional space, even when they are not sequentially adjacent. Furthermore, in the studied sequences, each with hundreds of residues, the number of functional nodes ranged between 1 and 30. These observations, firstly, highlight the need to consider subgraphs, rather than individual nodes, in protein graphs, as protein function is influenced by the local environment 
 and is usually carried by a cluster of residues, rather than isolated ones. It also suggests that aggressive pruning is necessary to accurately identify these few functional residues within graphs containing hundreds of nodes, necessitating a multi-stage pruning and refinement process. Finally, it is crucial that the pruning process preserves the subgraph structure, ensuring that the selected nodes form coherent clusters rather than being randomly scattered.
 

\subsection{Preliminaries}

Protein sequences are represented as graphs \( G(V, E) \), where the vertices \( V \) correspond to the protein residues, and the edges \( E \) represent the proximity of residues in three-dimensional space. The adjacency matrix \( A \in \mathbb{R}^{N \times N} \) for an \( N \)-residue protein graph is defined by calculating the contact map, where an edge is added between two nodes if the distance between their \( C_\alpha \) atoms is less than 10 \(\text{\AA}\). In this work, we use \( G(V, E) \) and \( G(Z, A) \) interchangeably, where \( V \) and \( E \) denote the set of vertices and edge list, while \( Z \in \mathbb{R}^{|V| \times D} \) and \( A \in \mathbb{R}^{|V| \times |V|} \) represent the node features and adjacency matrices, respectively, and $D$ is the chosen dimension for residue features. The goal of ProteinRPN is to predict a probability vector \( \hat{\mathbf{y}}_i^{(j)} \in \mathbb{R}^{l_j} \), where \( l_j \) denotes the number of GO terms associated with subontology \( j \in \{\text{BP}, \text{CC}, \text{MF}\}\). The vector \( \hat{\mathbf{y}}_i^{(j)} \) represents the predicted probabilities for the \( l_j \) GO terms, reflecting the likelihood of each protein being associated with multiple GO terms across all subontologies.

\subsubsection{Residue Features}
Residue features for $N$ nodes in any protein residue graph are derived through a two-step process. First, each node is assigned ESM-1b \cite{rives2021biological} embeddings $Z_E\in\mathbb{R}^{N\times D_E}$ to capture the intrinsic sequence-based information of the residues. 
In parallel, the residues are also label encoded according to their amino acid identities and transformed into embeddings $Z_R\in\mathbb{R}^{N\times D_R}$. These two feature sets are subsequently projected onto a common $D-$dimensional space and combined to form the final node embeddings $Z = Z_E + Z_R \in\mathbb{R}^{N \times D}$, effectively integrating both deep sequence information and basic residue identity.



\subsubsection{Enhanced Domain Knowledge}
To enhance the graph representation with domain-specific knowledge, we further extract the secondary structure of each residue for each protein using DSSP (Dictionary of Secondary Structure in Proteins \cite{dssp2015, Kabsch1983}, which is a database of secondary structure assignments for all protein entries in PDB \cite{berman2000protein}. Experimental evidence suggests that functional residues are more likely to be found in regions with defined secondary structures, such as alpha helices and beta sheets \cite{BARTLETT2002105}. To align the residue coordinate information with the secondary structure data, we perform sequence alignments between DSSP-processed variants and residues with available PDB coordinates, addressing any discrepancies that arise between these data sources.

\subsection{Functional Region Proposal Network}




Inspired by object detection models in computer vision, we propose a strategy analogous to region proposal networks in Faster R-CNN \cite{ren2015faster}, adapted for protein function prediction in graphs. By targeting regions containing functional residues, which are often a small subset of the protein, this approach improves functional understanding. To the best of our knowledge, this is the first work to introduce graph region proposals, applied specifically to protein function prediction.

The proposed Region Proposal Network employs $k$ layers of Graph Convolutional Networks (GCNs) \cite{kipf2017semisupervisedclassificationgraphconvolutional} to process protein graphs $G(Z, A)$. In particular, let $H^{(0)} = Z$ represent the initial hidden node embedding matrix. The hidden embeddings $H$ are updated iteratively as:

$$
H^{(i+1)} = \text{ReLU} \left( \tilde{D}^{-0.5} \tilde{A} \tilde{D}^{-0.5} H^{(i)} W^{(i)} \right)
$$
\noindent
where $\tilde{A} = A + I$ is the adjacency matrix with self-loops included, and $\tilde{D}$ is the diagonal degree matrix used for normalization. After $k$ message-passing layers, the final node embedding matrix $Z_1 = H^{(k)}\in\mathbb{R}^{N\times D_1}$ encapsulate information from their respective $k$-hop neighborhoods, effectively extending each node's receptive field to encompass its $k$-hop subgraph. Each node can, now, be designated as the representative of its corresponding $k$-hop subgraph, termed as an anchor. Consequently, this procedure transforms the original graph $G$ into a new graph $G'(Z_1, A)$, where each node in $G'$ corresponds to a subgraph in $G$. Empirical results indicate that setting $k=2$ is sufficient to capture functional residues within proteins.

The second step in the region proposal module involves localizing regions that are likely to contain functional residues. This is formulated as a node classification task, where the goal is to predict whether the anchor centred around each node contains a functionally relevant region. More precisely, given the node embeddings $Z_1$ after $k$ GCN layers, the classification of each node $v_i$ in the transformed graph $G'(Z_1, A)$ is performed using a Graph Attention Network (GAT) convolution \cite{veličković2018graphattentionnetworks}. The output for each node $v_i$ can be formulated as:

$$
\hat{y}_i = \sigma \left( \sum_{j \in \mathcal{N}(i)} \alpha_{ij} {W} H_j^{(k)} \right)
$$
\noindent
where, $\hat{y}_i$ is the predicted probability that the node $v_i$ in $G'$, which represents the $k$-hop subgraph $S_i$ in $G$, contains functional residues, $\alpha$ and $W$ are the attention scores and weight matrix, respectively, learnt by the GAT layer, $\mathcal{N}(i)$, represents the neighbors of node $i$ in the graph $G'(Z_1, A)$ 

Nodes predicted as functional are selected, and $k$-hop subgraphs (anchors) centered around these nodes are extracted. This results in a collection of anchors enriched for functional regions, ensuring high recall but with room for precision improvement. To address this, we introduce a pruning step that selectively retains the most functionally relevant subgraphs within the larger anchors. This pruning leverages a novel node-drop pooling layer that incorporates domain knowledge alongside a hierarchy-aware attention mechanism.  Rather than relying on conventional dot product attention, we utilize penumbral cone attention \cite{NEURIPS2023_a17251f8} for modeling the inherent hierarchical relationships in proteins. These hierarchies span multiple levels, from the arrangement of secondary and tertiary structures to the organization of functional domains and motifs, all the way up to the interactions of subunits within protein complexes. By capturing these complex dependencies, penumbral cone attention enables a more precise focus on critical functional regions within the protein graph, refining our predictions and improving precision without compromising recall.

Node drop pooling layers are commonly used to reduce graph size by selectively removing lower-scoring nodes while retaining higher-scoring ones based on their importance or features. Instead of removing low-scoring nodes, we evaluate the functionality of each node in the Node Drop Pooling layer. Subsequently, we leverage the Functional Attention Layer to enrich representations of high-scoring nodes, ensuring that lower-scoring nodes contributing valuable contextual information are retained to preserve the overall graph structure.

\subsubsection{Node Drop Pooling/Node Scoring} In order to obtain scores in our case, the feature embeddings extracted from these subgraphs are passed through GCN layers to obtain query and key representations. 

$$ q = \text{LeakyReLU}(GCN_1(G(Z_1, A))$$
$$ k = \text{LeakyReLU}(GCN_2(G(Z_1, A))$$

\noindent
These representations are then fed into a hierarchy-aware attention layer to decide which nodes to prune.

\subsubsection{Penumbral Cone Attention}
We employ cone attention \cite{gulcehre2018hyperbolic, NEURIPS2023_a17251f8} which serves as a seamless alternative to dot product attention, relying on hyperbolic entailment cones to model the hierarchies between the residue nodes. Specifically, we utilize hyperbolic distance attention, which
defines the similarity as $S\left(q_{i}, k_{i}\right)=\exp \left(-\beta d_{\mathbb{H}}\left(q_{i}, k_{i}\right)-c\right)$ where $d_{\mathbb{H}}$ is the hyperbolic distance. Following the previous work \cite{NEURIPS2023_a17251f8}, we use the Poincaré half-space model to calculate the hyperbolic distance.

\subsubsection{Proximity Scores}
To compute proximity scores, we first measure the pairwise distances between each residue and all other residues, akin to constructing a contact map. Rather than applying a threshold to these distances, the proximity score $P_i$ for residue $i$ is computed by summing the inverse distances between residue $i$ and all other residues $j$, i.e, $
P_i = \alpha_{ps} \sum_{j \neq i} \frac{1}{d_{ij}}
$, where $d_{ij}$ represents the distance between residues $i$ and $j$, and $\alpha_{ps}$ is a scaling factor that determines the influence of proximity on the final node score. This method prioritizes residues that are closely clustered with a few others, resulting in higher scores compared to residues that are moderately close to many others, aligning with our insights from PDBSite \cite{Ivanisenko2000PDBSite}.

\subsubsection{Secondary Structure Scores}
Certain functional residues have been observed to preferentially reside in regions of defined secondary structure. For instance, Bartlett et al. \cite{BARTLETT2002105} reports that catalytic residues are frequently located in alpha helices (39\%) and beta sheets (28\%), with a lower prevalence in loops and unstructured regions. To reflect this, we assign higher predicted scores to residues within alpha helices and beta sheets.

The final node scores, derived from the combination of the three components, are converted into probabilities using a sigmoid function. Residues with the highest probabilities are identified as functional for subsequent processing.

\subsection{Functional Attention Layer}

Once candidate functional residues are identified, their representations are refined through a functional attention layer. This layer assigns weights to edges based on their connectivity to predicted functional nodes, allowing the model to emphasize relationships critical to protein function. By incorporating multistage refinement, we iteratively enhance the accuracy of functional node identification. The edge-centric approach helps preserve the structural integrity of selected subgraphs, avoiding the fragmentation that can occur when individual nodes are selected in isolation, in line with insights from PDBSite.




We feed the original residue features $Z \in \mathbb{R}^{N \times D}$ as the node feature matrix for enrichment. For each edge $(i, j)$ in the graph, the model computes an attention score ${e}_{ij}$ using the concatenation of the feature vectors ${Z}_{i}$ and $Z_{j}$, followed by a learnable weight vector ${a} \in \mathbb{R}^{2D_1 \times 1}$ and a ReLU activation function, that would reduce all negative scores to zero, i.e., $
{e}_{ij} = \text{ReLU}\left({a}^\top [{Z}_{i} \, \Vert \, {Z}_{j}]\right) $. This attention score is then adjusted based on the node type ${z}_j \in \{0, 1\}$ of the target node $j$, modifying the score as follows:
\[
{e}_{ij} = \alpha_{FA} \cdot {e}_{ij} \cdot {z}_j + \beta_{FA} \cdot {e}_{ij} \cdot (1 - {z}_j) \cdot
\]
where $\alpha_{FA} \geq 1$ and $\beta_{FA} < 1$. This adjustment increases the attention for functional nodes while reducing it for contextual ones. For the purpose if this study, we use $\alpha_{FA}=1, \beta_{FA}=0.5$ in order to explicitly ensure focus on functional nodes. The attention coefficients $\alpha_{ij}$ are obtained by normalizing $e_{ij}$ across all neighbors. They determine how much influence a neighboring node $i$ has on the target node $j$. 

Finally, the updated feature vector for node $j$, ${Z}_{2j}$, is computed by aggregating the messages from its neighbors applying, weighted by the corresponding attention coefficients, i.e., ${Z_{2j}} = \sum_{i \in \mathcal{N}(j)}\alpha_{ij}\cdot W\cdot {Z}_{i}$, where the transformation matrix $W \in \mathbb{R}^{D_2\times D}$ is a learnable parameter as in the GAT layer and helps transform the initial extracted features of the nodes (residues) into a new space where relationships between residues can be more effectively captured. As a result of this operation, subgraphs surrounding functional residues—those likely to be critical for protein function—get more attention and influence the final node representations more significantly. 
 This approach enhances the model's ability to capture the rich, context-aware interactions between residues, leading to a more comprehensive understanding of the protein's functional regions.

\subsection{Graph Multiset Transformer}

In the final step, the enriched representations $Z_2$ are fed into a Graph Multiset Transformer (GMT) layer, which transforms node-level embeddings into a comprehensive graph-level representation by capturing both local interactions and global structure. The GMT layer introduces learnable super-nodes to capture long-distance structural information and aggregates this information into a unified graph representation.


\subsection{Optimization Framework}

Our model is optimized using a multi-component loss function that integrates cross-entropy loss \(\mathcal{L}_{\text{CE}}\) for multilabel classification, contrastive loss \(\mathcal{L}_{\text{con}}\), and a penalty term \(\mathcal{L}_{\text{penalty}}\) to minimize the number of disconnected components in the functional attention layer.




The contrastive loss, \(\mathcal{L}_{\text{con}}\), is a combination of supervised contrastive (SupCon) loss \cite{khosla2021supervisedcontrastivelearning} and self-supervised noise contrastive estimation (InfoNCE) loss \cite{oord2019representationlearningcontrastivepredictive}. SupCon encourages the model to cluster representations of proteins with similar GO terms, while InfoNCE ensures that the representations are robust to noise by maximizing the similarity between original and perturbed embeddings. The combined contrastive loss for a batch of $B$ proteins is defined as:

\begin{align*}
\mathcal{L}_{\text{con}} = \ &  \left( -\frac{1}{B} \sum_{i=1}^{B} \sum_{j \neq i} \mathbf{1}\{y_i \cap y_j \neq \emptyset\} \cdot \log \frac{\exp(\text{sim}(z_i, z_j)/\tau)}{\sum_{k \neq i} \exp(\text{sim}(z_i, z_k)/\tau)} \right)\\
&\cdot \alpha_{\text{SupCon}}  + \left( -\log \frac{\exp(\text{sim}(z_i, z_i')/\tau)}{\sum_{z_j} \exp(\text{sim}(z_i, z_j)/\tau)} \right)\alpha_{\text{NCE}}
\end{align*}

where \( z_i \) and \( z_j \) are the embeddings of proteins \( i \) and \( j \), \( \text{sim}(z_i, z_j) \) represents their cosine similarity, and \( \tau \) is a temperature parameter. The indicator function \( \mathbf{1}\{y_i \cap y_j \neq \emptyset\} \) ensures that only pairs with shared GO terms contribute to the SupCon loss, im order to adapt it to the multilabel case. The InfoNCE loss optimizes the similarity between the original and perturbed embeddings \( z_i \) and \( z_i' \). The hyperparameters \(\alpha_{\text{SupCon}}\) and \(\alpha_{\text{NCE}}\) control the contributions of the SupCon and InfoNCE losses.

To ensure that identified functional regions are structurally cohesive, reflecting the biological reality of interconnected functional residues, we introduce a connected components penalty. This discourages the formation of disconnected components in the functional attention layer and is defined as:

\[
\mathcal{L}_{\text{penalty}} = \alpha_{cc} \cdot \frac{\kappa(G''_i)}{\sum_{j=1}^M y_{ij}}
\]
\noindent
where \( G''_i = (V_f, E_f) \) denotes the subgraph of the \( i \)-th protein graph \( G_i \) induced by functional nodes, and \( \kappa(G''_i) \) represents the number of connected components in the subgraph. Finally, \(\alpha_{cc}\) is a hyperparameter that controls the strength of the penalty.

The overall loss function is formulated as:

\[
\mathcal{L} = \mathcal{L}_{\text{CE}} + \mathcal{L}_{\text{con}} + \mathcal{L}_{\text{penalty}}
\]

This comprehensive loss function guides the model toward producing biologically meaningful predictions by leveraging robust, context-aware graph representations while maintaining the structural coherence of the predicted functional subgraphs.

\section{Experiments}


In this section, we elabroate on the datasets, training setup, and evaluation criteria used for training the model.

\subsection{Datasets}
PDBSite \cite{Ivanisenko2000PDBSite} is a comprehensive dataset comprising biologically active sites derived from the Protein Data Bank (PDB) \cite{berman2000protein}. 
The dataset encompasses 4,723 active sites belonging to 197 different functions located within 603 proteins. PDBSite stands out among annotation databases due to its diverse representation of functional categories, enabling broad analysis across various protein functions. We leverage PDBSite to guide our model architecture and pretrain the model on predicting functional sites.



For protein function prediction, we utilize a dataset curated by \cite{gu2023hierarchical}, originally developed to train their model, HEAL, which serves as our baseline. This dataset is an adapted version of the DeepFRI dataset \cite{gligorijevic2021structure}, comprising 36,629 sequences sourced from the PDB database \cite{berman2000protein} and 42,994 from the SWISS-MODEL repository \cite{10.1093/nar/gkw1132}. Further details can be found in the Appendix.


\subsection{Experimental Setup}

We begin by training ProteinRPN on the PDBSite which is split into training and validation sets with an 80:20 ratio, with the goal of predicting all functional sites within a protein. The details of the pretraining can be found in the Appendix. Then we train the entire framework on the comprehensive protein function prediction task using the HEAL dataset.

We have conducted comprehensive experimennts to comapre ProteinRPN's [erformance to SOTA models. Those methods encompass sequence-based models such as BLAST \cite{altschul1990basic} and FunFams \cite{funfams}, sequence and PPI-based models like DeepGO \cite{deepgo2}, and sequence and structure-based models such as DeepFRI \cite{gligorijevic2021structure} and HEAL \cite{gu2023hierarchical}. 


We conduct ablation studies to assess the significance of each model component, including the impact of secondary structure, coordinate information, and contrastive learning losses. Additional studies to test the efficacy of other modules can be found in the Appendix.

Model predictions are evaluated using the standard Critical Assessment of Functional Annotation (CAFA) evaluator \cite{jiang2016expanded}. Protein-centric Fmax, the maximum F1 score over all prediction thresholds ranging from 0 to 1 with a step size of 0.1, is utilized. Smin, representing the semantic distance between predicted and actual annotations, considers the information content of each function. The function-centric AUPR is employed as a robust measure for situations with high class imbalance. Further details on the formulas and implementation are available in \cite{jiang2016expanded}, and comprehensive information on model training and hyperparameters can be found in the Appendix.

\section{Results and Analysis}

\begin{table*}[ht!]
\centering

\begin{tabular}{@{}lccc|ccc|ccc@{}}
\toprule
\textbf{Method} & \multicolumn{3}{c|}{\textbf{Fmax ($\uparrow$)}} & \multicolumn{3}{c|}{\textbf{AUPR ($\uparrow$)}} & \multicolumn{3}{c}{\textbf{Smin ($\downarrow$)}} \\
                & \textbf{BP} & \textbf{CC} & \textbf{MF} & \textbf{BP} & \textbf{CC} & \textbf{MF} & \textbf{BP} & \textbf{CC} & \textbf{MF} \\ 
\midrule
Blast           & 0.336       & 0.448       & 0.328       & 0.067       & 0.097       & 0.136       & 0.651       & 0.628       & 0.632       \\
FunFams         & 0.500       & 0.627       & 0.572       & 0.260       & 0.288       & 0.367       & 0.579       & 0.503       & 0.531       \\
DeepGO          & 0.493       & 0.594       & 0.577       & 0.182       & 0.263       & 0.391       & 0.577       & 0.550       & 0.472       \\
DeepFRI         & 0.540       & 0.613       & 0.625       & 0.261       & 0.274       & 0.495       & 0.543       & 0.527       & 0.437       \\
HEAL            & 0.581       & 0.673       & 0.708       & 0.298       & 0.415       & 0.630       & 0.504       & 0.462       & 0.369       \\
\textbf{ProteinRPN}      & \textbf{0.618}       & \textbf{0.691}       & \textbf{0.754}       & \textbf{0.344}       & \textbf{0.459}       & \textbf{0.683}       & \textbf{0.495}       & \textbf{0.458}       & \textbf{0.335}       \\ 
\bottomrule
\end{tabular}
\caption{Baseline Comparison: Fmax, AUPR, and Smin of different methods on the designated test set; best performances are highlighted in bold, i.e., for Fmax and AUPR, we consider the highest, while for Smin we consider the lowest value}
\label{tab:res}
\end{table*}

\subsection{GO term Prediction}

Table \ref{tab:res} presents the performance metrics of ProteinRPN in comparison to all baseline models on the HEAL dataset. ProteinRPN consistently outperforms the baselines across all metrics, showing notable improvements over the HEAL model. Specifically, ProteinRPN achieves higher Fmax scores, with gains of 6.4\% in Biological Process (BP), 2.7\% in Cellular Component (CC), and 7.1\% in Molecular Function (MF) ontologies. Beyond Fmax, ProteinRPN also demonstrates superior performance in Smin and Area Under the Precision-Recall Curve (AUPR), highlighting its effectiveness in predicting protein function GO terms.

Moreover, as shown in Table \ref{tab:ablation}, the ablation study reveals that both contrastive learning and the incorporation of domain knowledge positively contribute to the model’s overall performance.

During pretraining, the region proposal module exhibits strong performance, achieving an ROC of 0.95 in the anchor functionality prediction task and 0.85 in the pruning task. Although direct comparison is limited due to the absence of established baselines, the module's effectiveness is evident in downstream functional prediction tasks.

Overall, the results demonstrate that enabling the model to detect and focus on residues critical for function significantly enhances its performance. This improvement is primarily driven by the multistage refinement approach, which efficiently localizes functional residues within protein structures.

\begin{table*}[ht!]
\centering
\begin{tabular}{@{}lccc|ccc|ccc@{}}
\toprule
\textbf{Model} & \multicolumn{3}{c|}{\textbf{Fmax ($\uparrow$)}} & \multicolumn{3}{c|}{\textbf{AUPR ($\uparrow$)}} & \multicolumn{3}{c}{\textbf{Smin ($\downarrow$)}} \\
               & \textbf{BP} & \textbf{CC} & \textbf{MF} & \textbf{BP} & \textbf{CC} & \textbf{MF} & \textbf{BP} & \textbf{CC} & \textbf{MF} \\ 
\midrule
ProteinRPN CL          & \textbf{0.6175} & \textbf{0.6906} & \textbf{0.7542} & \textbf{0.3438} & 0.4527 & 0.6833 & \textbf{0.4948} & \textbf{0.4576} & \textbf{0.3350} \\
ProteinRPN w/o CL      & 0.6009 & 0.6878 & 0.7408 & 0.3223 & 0.4166 & 0.6479 & 0.5062 & 0.4587 & 0.3557 \\
ProteinRPN w/o SS w CL & 0.6114 & 0.6894 & 0.7498 & 0.3426 & \textbf{0.4591} & 0.6778 & 0.4984 & \textbf{0.4576} & 0.3421 \\
ProteinRPN w/o SS w/o CL & 0.5975 & 0.6801 & 0.7364 & 0.3161 & 0.4242 & 0.6446 & 0.5088 & 0.4674 & 0.3547 \\
\bottomrule
\end{tabular}
\caption{Ablation Studies: Fmax, AUPR, and Smin of different variants of ProteinRPN, where CL: Contrastive Learning, SS: secondary structure and proximity scoring; best performances are highlighted in bold, i.e., for Fmax and AUPR, we consider the highest, while for Smin we consider the lowest value. On removing the Contrastive Learning module, there is a moderate decrease in performance across all three GO domains; further, removing Domain Knowledge from the node pooling layer is also seen to impact performance negatively.}
\label{tab:ablation}
\end{table*}

\begin{figure}[ht!]
    \centering
    \includegraphics[width=0.49\linewidth]{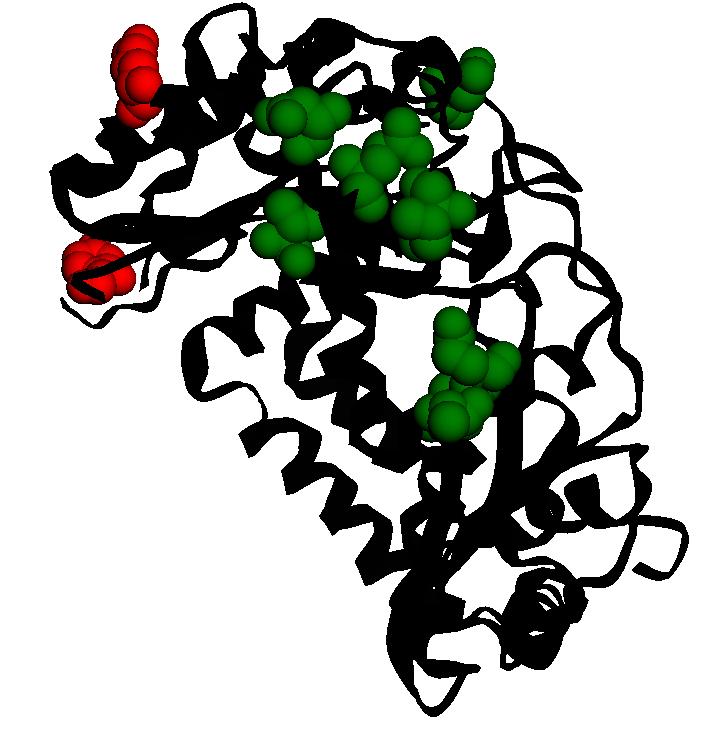}
    \includegraphics[width=0.49\linewidth]{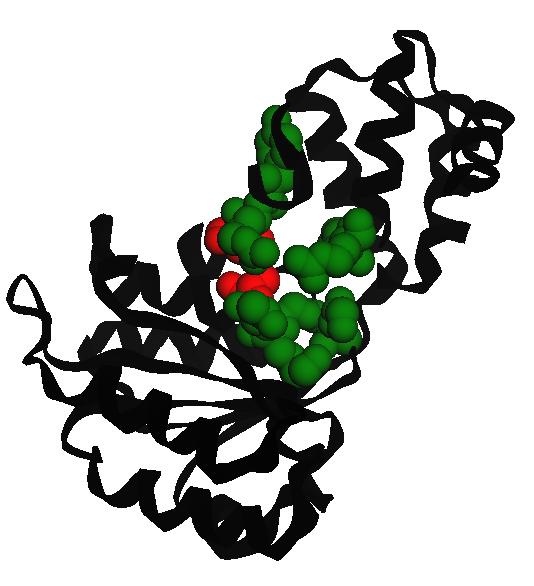}
    \caption{Visual Demonstration of Region Proposal Network detected residues in proteins (a) 2BCC-B and (b) 2CHG-A}
    
    \label{fig:protres}
\end{figure}

\subsection{Functional Residue Visualization}
We evaluate the functional residue predictor in ProteinRPN by analyzing specific proteins. For example, on protein 2BCC (B chain, 422 residues, 10 functional), ProteinRPN accurately identifies 8 functional residues, with region proposals covering subgraphs of 28 residues, as shown in Fig. \ref{fig:protres}(a). Functional residues predicted correctly are highlighted in green, while missed ones are marked in red.

Similarly, for protein 2CHG (A chain, 226 residues, 11 functional), the model successfully identifies 9 functional residues, with region proposals covering 43 residues. As shown in Fig. \ref{fig:protres}(b), the correctly identified residues are closely clustered within the structure, while the missed residues are located farther from the cluster. These results demonstrate its ability to accurately identify and localize constellations of functional residues.

\section{Conclusion}
In this work, we introduced ProteinRPN, a novel graph-based model equipped with graph region proposal networks which is designed to identify and refine functional regions within protein residue graphs. By leveraging hierarchical attention mechanisms, domain-specific knowledge, and multistage refinement, through a combination of supervised contrastive learning and self-supervised InfoNCE loss, ProteinRPN significantly improves the accuracy of protein function prediction across GO terms. Our results demonstrate substantial gains over SOTA methods, with enhanced precision in identifying functional residues and preserving structural integrity in predicted subgraphs. 

While our model provides generalized insights across a range of protein functions, the current analysis is based on a limited set of protein structures. Future work will focus on extending the model's capabilities by incorporating diverse knowledge sources and exploring additional mechanisms to further enhance the accuracy and scalability of protein function prediction.


\bibliography{aaai25.bib}

\begin{thebibliography}{40}
\providecommand{\natexlab}[1]{#1}

\bibitem[{gen(2021)}]{gene2021gene}
 2021.
\newblock The Gene Ontology resource: enriching a GOld mine.
\newblock \emph{Nucleic acids research}, 49(D1): D325--D334.

\bibitem[{uni(2023)}]{uniprot2023uniprot}
 2023.
\newblock UniProt: the universal protein knowledgebase in 2023.
\newblock \emph{Nucleic acids research}, 51(D1): D523--D531.

\bibitem[{Aleksander et~al.(2023)Aleksander, Balhoff, Carbon, Cherry, Drabkin, Ebert, Feuermann, Gaudet, Harris et~al.}]{gene2023gene}
Aleksander, S.~A.; Balhoff, J.; Carbon, S.; Cherry, J.~M.; Drabkin, H.~J.; Ebert, D.; Feuermann, M.; Gaudet, P.; Harris, N.~L.; et~al. 2023.
\newblock The gene ontology knowledgebase in 2023.
\newblock \emph{Genetics}, 224(1): iyad031.

\bibitem[{Altschul et~al.(1990)Altschul, Gish, Miller, Myers, and Lipman}]{altschul1990basic}
Altschul, S.~F.; Gish, W.; Miller, W.; Myers, E.~W.; and Lipman, D.~J. 1990.
\newblock Basic local alignment search tool.
\newblock \emph{Journal of Molecular Biology}, 215(3): 403--410.

\bibitem[{Bartlett et~al.(2002)Bartlett, Porter, Borkakoti, and Thornton}]{BARTLETT2002105}
Bartlett, G.~J.; Porter, C.~T.; Borkakoti, N.; and Thornton, J.~M. 2002.
\newblock Analysis of Catalytic Residues in Enzyme Active Sites.
\newblock \emph{Journal of Molecular Biology}, 324(1): 105--121.

\bibitem[{Berman et~al.(2000)Berman, Westbrook, Feng, Gilliland, Bhat, Weissig, Shindyalov, and Bourne}]{berman2000protein}
Berman, H.~M.; Westbrook, J.; Feng, Z.; Gilliland, G.; Bhat, T.~N.; Weissig, H.; Shindyalov, I.~N.; and Bourne, P.~E. 2000.
\newblock The Protein Data Bank.
\newblock \emph{Nucleic Acids Research}, 28(1): 235--242.

\bibitem[{Bienert et~al.(2016)Bienert, Waterhouse, deBeer, Tauriello, Studer, Bordoli, and Schwede}]{10.1093/nar/gkw1132}
Bienert, S.; Waterhouse, A.; deBeer, T.~A.; Tauriello, G.; Studer, G.; Bordoli, L.; and Schwede, T. 2016.
\newblock {The SWISS-MODEL Repository—new features and functionality}.
\newblock \emph{Nucleic Acids Research}, 45(D1): D313--D319.

\bibitem[{Cho, Berger, and Peng(2016)}]{cho2016compact}
Cho, H.; Berger, B.; and Peng, J. 2016.
\newblock Compact integration of multi-network topology for functional analysis of genes.
\newblock \emph{Cell systems}, 3(6): 540--548.

\bibitem[{Das et~al.(2015)Das, Lee, Sillitoe, Dawson, Lees, and Orengo}]{funfams}
Das, S.; Lee, D.; Sillitoe, I.; Dawson, N.~L.; Lees, J.~G.; and Orengo, C.~A. 2015.
\newblock {Functional classification of CATH superfamilies: a domain-based approach for protein function annotation}.
\newblock \emph{Bioinformatics}, 31(21): 3460--3467.

\bibitem[{Fa et~al.(2018)Fa, Cozzetto, Wan, and Jones}]{fa2018predicting}
Fa, R.; Cozzetto, D.; Wan, C.; and Jones, D.~T. 2018.
\newblock Predicting human protein function with multi-task deep neural networks.
\newblock \emph{PloS one}, 13(6): e0198216.

\bibitem[{Gligorijevi{\'c} et~al.(2021)Gligorijevi{\'c}, Renfrew, Kosciolek, Leman, Berenberg, Vatanen, Chandler, Taylor, Fisk, Vlamakis et~al.}]{gligorijevic2021structure}
Gligorijevi{\'c}, V.; Renfrew, P.~D.; Kosciolek, T.; Leman, J.~K.; Berenberg, D.; Vatanen, T.; Chandler, C.; Taylor, B.~C.; Fisk, I.~M.; Vlamakis, H.; et~al. 2021.
\newblock Structure-based protein function prediction using graph convolutional networks.
\newblock \emph{Nature communications}, 12(1): 3168.

\bibitem[{Gong, Ning, and Tian(2016)}]{gong2016gofdr}
Gong, Q.; Ning, W.; and Tian, W. 2016.
\newblock GoFDR: a sequence alignment based method for predicting protein functions.
\newblock \emph{Methods}, 93: 3--14.

\bibitem[{Gu et~al.(2023)Gu, Luo, Chen, Deng, and Lai}]{gu2023hierarchical}
Gu, Z.; Luo, X.; Chen, J.; Deng, M.; and Lai, L. 2023.
\newblock Hierarchical graph transformer with contrastive learning for protein function prediction.
\newblock \emph{Bioinformatics}, 39(7): btad410.

\bibitem[{Gulcehre et~al.(2018)Gulcehre, Denil, Malinowski, Razavi, Pascanu, Hermann, Battaglia, Bapst, Raposo, Santoro et~al.}]{gulcehre2018hyperbolic}
Gulcehre, C.; Denil, M.; Malinowski, M.; Razavi, A.; Pascanu, R.; Hermann, K.~M.; Battaglia, P.; Bapst, V.; Raposo, D.; Santoro, A.; et~al. 2018.
\newblock Hyperbolic attention networks.
\newblock \emph{arXiv preprint arXiv:1805.09786}.

\bibitem[{Ivanisenko, Grigorovich, and Kolchanov(2000)}]{Ivanisenko2000PDBSite}
Ivanisenko, V.; Grigorovich, D.; and Kolchanov, N. 2000.
\newblock PDBSite: a database on biologically active sites and their spatial surroundings in proteins with known tertiary structure.
\newblock In \emph{The Second International Conference on Bioinformatics of Genome Regulation and Structure (BGRS'2000)}, volume~2, 171--174. Novosibirsk, Russia.

\bibitem[{Jeffery(2023)}]{challengeCJ2023}
Jeffery, C.~J. 2023.
\newblock Current successes and remaining challenges in protein function prediction.
\newblock \emph{Frontiers in Bioinformatics}, 3.

\bibitem[{Jiang et~al.(2016)Jiang, Oron, Clark et~al.}]{jiang2016expanded}
Jiang, Y.; Oron, T.~R.; Clark, W.~T.; et~al. 2016.
\newblock An expanded evaluation of protein function prediction methods shows an improvement in accuracy.
\newblock \emph{Genome Biology}, 17(1): 184.

\bibitem[{Kabsch and Sander(1983)}]{Kabsch1983}
Kabsch, W.; and Sander, C. 1983.
\newblock Dictionary of protein secondary structure: pattern recognition of hydrogen-bonded and geometrical features.
\newblock \emph{Biopolymers}, 22: 2577--2637.

\bibitem[{Khosla et~al.(2021)Khosla, Teterwak, Wang, Sarna, Tian, Isola, Maschinot, Liu, and Krishnan}]{khosla2021supervisedcontrastivelearning}
Khosla, P.; Teterwak, P.; Wang, C.; Sarna, A.; Tian, Y.; Isola, P.; Maschinot, A.; Liu, C.; and Krishnan, D. 2021.
\newblock Supervised Contrastive Learning.
\newblock arXiv:2004.11362.

\bibitem[{Kipf and Welling(2017)}]{kipf2017semisupervisedclassificationgraphconvolutional}
Kipf, T.~N.; and Welling, M. 2017.
\newblock Semi-Supervised Classification with Graph Convolutional Networks.
\newblock arXiv:1609.02907.

\bibitem[{Kulmanov et~al.(2024)Kulmanov, Guzm{\'a}n-Vega, Duek~Roggli, Lane, Arold, and Hoehndorf}]{kulmanov2024protein}
Kulmanov, M.; Guzm{\'a}n-Vega, F.~J.; Duek~Roggli, P.; Lane, L.; Arold, S.~T.; and Hoehndorf, R. 2024.
\newblock Protein function prediction as approximate semantic entailment.
\newblock \emph{Nature Machine Intelligence}, 6(2): 220--228.

\bibitem[{Kulmanov and Hoehndorf(2022)}]{kulmanov2022deepgozero}
Kulmanov, M.; and Hoehndorf, R. 2022.
\newblock DeepGOZero: improving protein function prediction from sequence and zero-shot learning based on ontology axioms.
\newblock \emph{Bioinformatics}, 38(Supplement\_1): i238--i245.

\bibitem[{Kulmanov, Khan, and Hoehndorf(2017)}]{deepgo2}
Kulmanov, M.; Khan, M.~A.; and Hoehndorf, R. 2017.
\newblock {DeepGO: predicting protein functions from sequence and interactions using a deep ontology-aware classifier}.
\newblock \emph{Bioinformatics}, 34(4): 660--668.

\bibitem[{Kulmanov, Khan, and Hoehndorf(2018)}]{kulmanov2018deepgo}
Kulmanov, M.; Khan, M.~A.; and Hoehndorf, R. 2018.
\newblock DeepGO: predicting protein functions from sequence and interactions using a deep ontology-aware classifier.
\newblock \emph{Bioinformatics}, 34(4): 660--668.

\bibitem[{Lai and Xu(2022)}]{lai2022accurate}
Lai, B.; and Xu, J. 2022.
\newblock Accurate protein function prediction via graph attention networks with predicted structure information.
\newblock \emph{Briefings in Bioinformatics}, 23(1): bbab502.

\bibitem[{Li(2011)}]{li2011short}
Li, H. 2011.
\newblock A short introduction to learning to rank.
\newblock \emph{IEICE TRANSACTIONS on Information and Systems}, 94(10): 1854--1862.

\bibitem[{Pan et~al.(2023)Pan, Li, Bi, Wang, Gasser, Purcell, Akutsu, Webb, Imoto, and Song}]{pan2023pfresgo}
Pan, T.; Li, C.; Bi, Y.; Wang, Z.; Gasser, R.~B.; Purcell, A.~W.; Akutsu, T.; Webb, G.~I.; Imoto, S.; and Song, J. 2023.
\newblock PFresGO: an attention mechanism-based deep-learning approach for protein annotation by integrating gene ontology inter-relationships.
\newblock \emph{Bioinformatics}, 39(3): btad094.

\bibitem[{Ren et~al.(2015)Ren, He, Girshick, and Sun}]{ren2015faster}
Ren, S.; He, K.; Girshick, R.; and Sun, J. 2015.
\newblock Faster r-cnn: Towards real-time object detection with region proposal networks.
\newblock \emph{Advances in neural information processing systems}, 28.

\bibitem[{Rives et~al.(2021)Rives, Meier, Sercu, Goyal, Lin, Liu, Guo, Ott, Zitnick, Ma et~al.}]{rives2021biological}
Rives, A.; Meier, J.; Sercu, T.; Goyal, S.; Lin, Z.; Liu, J.; Guo, D.; Ott, M.; Zitnick, C.~L.; Ma, J.; et~al. 2021.
\newblock Biological structure and function emerge from scaling unsupervised learning to 250 million protein sequences.
\newblock \emph{Proceedings of the National Academy of Sciences}, 118(15): e2016239118.

\bibitem[{Tang et~al.(2018)Tang, Wang, Wang, Yan, Liu, Huang, and Yuille}]{tang2018weakly}
Tang, P.; Wang, X.; Wang, A.; Yan, Y.; Liu, W.; Huang, J.; and Yuille, A. 2018.
\newblock Weakly supervised region proposal network and object detection.
\newblock In \emph{Proceedings of the European conference on computer vision (ECCV)}, 352--368.

\bibitem[{Tian, Arakaki, and Skolnick(2004)}]{tian2004eficaz}
Tian, W.; Arakaki, A.~K.; and Skolnick, J. 2004.
\newblock EFICAz: a comprehensive approach for accurate genome-scale enzyme function inference.
\newblock \emph{Nucleic acids research}, 32(21): 6226--6239.

\bibitem[{Touw et~al.(2015)Touw, Baakman, Black, te~Beek, Krieger, Joosten, and Vriend}]{dssp2015}
Touw, W.~G.; Baakman, C.; Black, J.; te~Beek, T. A.~H.; Krieger, E.; Joosten, R.~P.; and Vriend, G. 2015.
\newblock A series of PDB related databases for everyday needs.
\newblock \emph{Nucleic Acids Research}, 43(Database issue): D364--D368.

\bibitem[{Tseng et~al.(2023)Tseng, Yu, Liu, and De~Sa}]{NEURIPS2023_a17251f8}
Tseng, A.; Yu, T.; Liu, T.; and De~Sa, C.~M. 2023.
\newblock Coneheads: Hierarchy Aware Attention.
\newblock In Oh, A.; Naumann, T.; Globerson, A.; Saenko, K.; Hardt, M.; and Levine, S., eds., \emph{Advances in Neural Information Processing Systems}, volume~36, 51421--51433. Curran Associates, Inc.

\bibitem[{van~den Oord, Li, and Vinyals(2019)}]{oord2019representationlearningcontrastivepredictive}
van~den Oord, A.; Li, Y.; and Vinyals, O. 2019.
\newblock Representation Learning with Contrastive Predictive Coding.
\newblock arXiv:1807.03748.

\bibitem[{Veličković et~al.(2018)Veličković, Cucurull, Casanova, Romero, Liò, and Bengio}]{veličković2018graphattentionnetworks}
Veličković, P.; Cucurull, G.; Casanova, A.; Romero, A.; Liò, P.; and Bengio, Y. 2018.
\newblock Graph Attention Networks.
\newblock arXiv:1710.10903.

\bibitem[{Wang et~al.(2023)Wang, You, Liu, Xiong, and Zhu}]{wang2023netgo}
Wang, S.; You, R.; Liu, Y.; Xiong, Y.; and Zhu, S. 2023.
\newblock NetGO 3.0: protein language model improves large-scale functional annotations.
\newblock \emph{Genomics, Proteomics \& Bioinformatics}, 21(2): 349--358.

\bibitem[{You et~al.(2021)You, Yao, Mamitsuka, and Zhu}]{you2021deepgraphgo}
You, R.; Yao, S.; Mamitsuka, H.; and Zhu, S. 2021.
\newblock DeepGraphGO: graph neural network for large-scale, multispecies protein function prediction.
\newblock \emph{Bioinformatics}, 37(Supplement\_1): i262--i271.

\bibitem[{You et~al.(2019)You, Yao, Xiong, Huang, Sun, Mamitsuka, and Zhu}]{you2019netgo}
You, R.; Yao, S.; Xiong, Y.; Huang, X.; Sun, F.; Mamitsuka, H.; and Zhu, S. 2019.
\newblock NetGO: improving large-scale protein function prediction with massive network information.
\newblock \emph{Nucleic acids research}, 47(W1): W379--W387.

\bibitem[{You et~al.(2018)You, Zhang, Xiong, Sun, Mamitsuka, and Zhu}]{GOLaberler2018}
You, R.; Zhang, Z.; Xiong, Y.; Sun, F.; Mamitsuka, H.; and Zhu, S. 2018.
\newblock {GOLabeler: improving sequence-based large-scale protein function prediction by learning to rank}.
\newblock \emph{Bioinformatics}, 34(14): 2465--2473.

\bibitem[{Zhao, Liu, and Wang(2022)}]{zhao2022panda2}
Zhao, C.; Liu, T.; and Wang, Z. 2022.
\newblock PANDA2: protein function prediction using graph neural networks.
\newblock \emph{NAR genomics and bioinformatics}, 4(1): lqac004.

\end{thebibliography}
\appendix

\end{document}


\maketitle

\section{Appendix}

\subsection{Correction}
In the Functional Attention Layer section, we previously stated, ``\ldots the model computes an attention score ${e}_{ij}$ using the concatenation of the feature vectors ${Z}_{1i}$ and ${Z}_{1j}$ \ldots” This was a typographical error. The correct formulation should use ${Z}_{i}$ and ${Z}_{j}$, respectively, as the original feature vectors are being concatenated.



\subsection{Protein Function Prediction Dataset}

For protein function prediction, we utilize a dataset curated by Gu et al. \cite{gu2023hierarchical}, originally developed for training their HEAL model, which serves as our baseline. This dataset is a modified version of the DeepFRI dataset \cite{gligorijevic2021structure}, containing sequences sourced from the PDB database and the SWISS-MODEL repository. Each dataset entry includes contact maps, where residues are considered in contact if the distance between their $C_\alpha$ atoms is less than 10 \(\text{\AA}\), and ESM-1b embeddings for individual residues. The dataset is split into training, validation, and test sets, maintaining an 8:1:1 ratio as can be seen in Table \ref{tab:seq}.

For predicting Gene Ontology (GO) terms in alignment with standard practices, we use the leaf nodes of the directed acyclic graph as prediction labels, resulting in 489 binary prediction tasks for Molecular Function (GO-MF), 1,943 for Biological Process (GO-BP), and 320 for Cellular Component (GO-CC).

We employ a multi-cutoff split method, as utilized by Gligorijevic et al. (2021), ensuring that the test set comprises only PDB chains with sequence identities of 95\% or less. Furthermore, each PDB chain in the test set is guaranteed to have at least one experimentally validated GO term for each GO domain. The test set remains consistent with that used by DeepFRI and other baseline models to maintain parity of comparison.

\begin{table}[ht]
\centering

\begin{tabular}{@{}lccc@{}}
\toprule
\textbf{Dataset}    & \textbf{Train} & \textbf{Val} & \textbf{Test} \\
\midrule
PDB                 & 29,893         & 3,322        & 3,414         \\
SWISS-MODEL       & 38,185         & 4,242        & 567           \\
\bottomrule
\end{tabular}
\caption{Dataset distribution for training, validation, and testing.}
\label{tab:seq}
\end{table}

In order to train the ProteinRPN model, we combine the training and validation sequences from the PDB and SWISS-MODEL repositories. To maintain consistency with the DeepFRI benchmark dataset (Gligorijevic et al., 2021), we use only the PDB test set as our test set, while the SWISS-MODEL test set is incorporated into the training set.

\subsection{Region Proposal Module Pretraining}

In alignment with the region proposal training procedure for downstream tasks, we treat each node's $k$-hop subgraph as an anchor, where the node itself serves as the representative of the anchor. This reformulates the original graph $G$ into a transformed graph $G'(H', A)$, where each node in $G'$ corresponds to a subgraph in $G$. Two Graph Neural Network (GNN) layers are applied to $G'$. The first layer predicts the likelihood of each anchor containing a functional site, while the second generates a vector determining which nodes within the anchor’s subgraph should be retained or pruned.

Anchors are labeled as positive based on their Jaccard Similarity overlap with ground-truth annotations. Specifically, an anchor is considered positive if its Jaccard Similarity exceeds a threshold of 0.7, and negative if it falls below 0.3. This dual-threshold strategy ensures a clear separation between functionally relevant and irrelevant regions.

The optimization of ProteinRPN is guided by a loss function that combines classification and regression tasks, inspired by Faster R-CNN:
\[
\mathcal{L}(\{p_i\}, \{t_i\}) = \frac{1}{N_{\text{cls}}} \sum_{i} \mathcal{L}_{\text{cls}}(p_i, p_i^*) + \lambda \frac{1}{N_{\text{reg}}} \sum_{i} p_i^* \mathcal{L}_{\text{pruning}}(G_i, G_i^*)
\]
where:
\begin{itemize}
    \item $\mathcal{L}_{\text{cls}}(p_i, p_i^*)$ is the classification loss, computed as binary cross-entropy between the predicted probability $p_i$ and the ground truth $p_i^*$, indicating the presence of a functional site.
    \item $\mathcal{L}_{\text{pruning}}(G_i, G_i^*)$ measures pruning accuracy, defined by the discrepancy in overlap between pruned and ground-truth subgraphs.
    \item $\lambda$ is a balancing factor regulating the relative contributions of classification and pruning losses. Through experimentation, $\lambda = 1$ was found to offer an optimal tradeoff between identifying functional sites and accurately delineating their spatial boundaries.
\end{itemize}

This structured loss function, inspired by that of Faster R-CNN \cite{ren2015faster}, facilitates precise refinement of functional site predictions and their boundaries within the protein structure. High-confidence anchors and their associated subgraphs are aggregated, with overlapping subgraphs being unified to produce comprehensive functional annotations. Specifically, anchors with a high probability of containing functional regions (probability of functionality $>$ 0.7) are identified. Within these anchors, residues with high retention probabilities (probability of retention $>$ 0.5) are retained as functional residues. Finally, functional residues from nearby high-confidence anchors are unified.

\subsection{Efficacy of Contrastive Learning and Graph Multiset Transformer Modules}
To supplement our ablation studies and assess the generalizability of our model, we investigate the efficacy of individual components. Specifically, we explore various contrastive learning objectives, such as SINCERE, and substitute the Graph Multiset Transformer module with other variants of graph transformers to evaluate their impact on model performance.

\subsubsection{SINCERE Loss}

For our contrastive learning objective, we employed a combination of supervised contrastive loss (SupCon) and InfoNCE. Recent literature suggests advanced formulations that integrate both losses in a unified framework. One such formulation is the Supervised Information Noise-Contrastive Estimation (SINCERE) loss \cite{feeney2024sinceresupervisedinformationnoisecontrastive}. SINCERE is an extension of InfoNCE, specifically adapted for supervised learning, and addresses the issue of within-class repulsion seen in SupCon by maximizing within-class similarity.

For each protein graph $G$, we generate multiple views by perturbing node embeddings with random noise. SINCERE aims to maximize the similarity between views of the same protein while minimizing similarity across different proteins. Mathematically, the loss is defined as:
\[
\mathcal{L}_{\text{SINCERE}} = -\frac{1}{M} \sum_{m=1}^{M} \log \frac{\exp(\text{sim}(\mathbf{z}_m, \mathbf{z}_m') / \tau)}{\sum_{m'=1}^{M} \exp(\text{sim}(\mathbf{z}_m, \mathbf{z}_{m'}) / \tau)}
\]
where $\text{sim}(\mathbf{z}_m, \mathbf{z}_m')$ represents the cosine similarity between two augmented views of the $m$-th protein graph, and $\tau$ is the temperature parameter set to 0.1. This contrastive framework enhances the model’s ability to learn robust and discriminative graph representations by promoting similarity for functionally related regions while ensuring distinct representations for different proteins.

We replace our usual contrastive loss with SINCERE and combine it with binary cross-entropy (BCE) loss for classification. However, our experiments reveal that the combination of SupCon and InfoNCE outperforms using either InfoNCE alone or SINCERE as can be seen in Table \ref{tab:supres}. This is likely because the external combination of SupCon and InfoNCE offers more fine-grained control over their relative importance, while SINCERE inherently predefines their integration.

\subsubsection{Polynormer: A Polynomial-Expressive Graph Transformer}

In a final experiment, we assess the efficacy of the Graph Multiset Transformer (GMT) module by substituting it with an empirically effective graph transformer, Polynormer \cite{deng2024polynormer}.

Polynormer \cite{deng2024polynormer} is designed to balance expressivity and scalability in graph learning tasks. Traditional GNNs often suffer from over-smoothing and limited expressive power when modeling complex functions. Polynormer addresses these issues by learning high-degree polynomials on graph data, enabling it to capture intricate node relationships while maintaining linear computational complexity. Formally, node representations at layer $l$ are computed as:

\[
X^{(l)} = \left(W^{(l)} X^{(l-1)}\right) \odot \left(X^{(l-1)} + B^{(l)}\right)
\]

where $W^{(l)} \in \mathbb{R}^{n \times n}$ and $B^{(l)} \in \mathbb{R}^{n \times d}$ are trainable weight matrices, $\odot$ denotes the Hadamard product, and $X^{(l-1)}$ is the input node feature matrix at layer $l-1$. Polynormer models polynomial functions where the degree of the polynomial grows exponentially with the number of layers, allowing a depth $L$ Polynormer to represent polynomials of degree $2^L$.

Polynormer integrates graph structure through two types of equivariant attention mechanisms: local attention, which incorporates adjacency information, and global attention, which captures higher-order interactions across the entire graph. This local-to-global attention mechanism mirrors the intuition behind using GMT, making Polynormer a suitable candidate for comparison.

However, in Table \ref{tab:supres} we observe a significant performance drop when substituting GMT with Polynormer. This highlights the effectiveness of GMT’s supernode-based topological pooling over traditional pooling approaches that treat nodes equally. The introduction of supernode representations in GMT proves to be more adept at capturing key functional substructures, which are critical for accurate function prediction.



\begin{table*}[ht!]
\centering

\begin{tabular}{@{}lccc|ccc|ccc@{}}
\toprule
\textbf{Model} & \multicolumn{3}{c|}{\textbf{Fmax ($\uparrow$)}} & \multicolumn{3}{c|}{\textbf{AUPR ($\uparrow$)}} & \multicolumn{3}{c}{\textbf{Smin ($\downarrow$)}} \\
               & \textbf{BP} & \textbf{CC} & \textbf{MF} & \textbf{BP} & \textbf{CC} & \textbf{MF} & \textbf{BP} & \textbf{CC} & \textbf{MF} \\ 
\midrule
ProteinRPN CL          & \textbf{0.6175} & \textbf{0.6906} & \textbf{0.7542} & \textbf{0.3438} & \textbf{0.4527} & \textbf{0.6833} & \textbf{0.4948} & \textbf{0.4576} & \textbf{0.3350} \\
ProteinRPN w SINCERE   & 0.5823 & 0.6676 & 0.7513 & 0.3128 & 0.3990 & \textbf{0.6833} & 0.5180 & 0.4779 & 0.3461 \\
ProteinRPN w Polynormer & 0.5102 & 0.6269 & 0.5877 & 0.2012 & 0.3076 & 0.4030 & 0.5629 & 0.5257 & 0.4908 \\
\bottomrule
\end{tabular}
\caption{Fmax, AUPR, and Smin of different variants of the proposed model. Best performance in bold where applicable.}
\label{tab:supres}
\end{table*}

\subsection{Training Setup}

We train the proposed ProteinRPN model using the Adam optimizer with a learning rate of 0.0001 and a batch size of 48 for 100 epochs. All models are implemented using PyTorch and the PyTorch Geometric library \cite{paszke2017automatic, Fey/Lenssen/2019}. Training is conducted on a single NVIDIA A100 80 GB Tensor Core GPU, with training times of approximately 10 hours per model using a batch size of 48.

\noindent
\textbf{Hyperparameter Settings:} All hyperparameters are tuned using Optuna \cite{optuna}, which employs Tree-structured Parzen Estimator (TPE) sampling. The input feature dimension is set to $D = 1280$, and the hidden channels in the $k$-layer GCN are $D_1 = 256$ with $k = 2$. The output dimension for the functional attention layer is $D_2 = 512$. The loss function tuning parameters are set to $\alpha_{\text{cc}} = 0.001$, $\alpha_{\text{SupCon}} = 0.01$, and $\alpha_{\text{NCE}} = 0.01$.

\subsection{Code and Data Availability}
All the code and datasets have been made available here: \url{https://drive.google.com/drive/folders/18s2baI9Nt0ztgzms_dGMGOsP5GvndvPl?usp=sharing}.

\bibliography{LaTeX/aaai25}